\documentclass[pre,twocolumn,superscriptaddress,amsmath,amssymb,floatfix,showpacs]{revtex4}
\usepackage{graphicx}
\usepackage{epsfig}
\usepackage{dcolumn}
\usepackage{bm}
\usepackage{psfrag}

\usepackage[normalem]{ulem}
\usepackage{color}
\usepackage{soul}

\newcommand{\nvec}{\mathbf{n}}

\newcommand{\Qij}{Q_{ij}}

\begin{document}

\bibliographystyle{apsrev}

\title  {Key-Lock Colloids in a Nematic Liquid Crystal}

\author{Nuno M. Silvestre}
\email[]{nmsilvestre@ciencias.ulisboa.pt}
\affiliation{Departamento de F{\'\i}sica da Faculdade de Ci\^encias,
Universidade de Lisboa, Campo Grande, P-1649-003 Lisboa, Portugal}
\affiliation{Centro de F{\'\i}sica Te\'orica e Computacional,
Universidade de Lisboa, Campo Grande, P-1649-003 Lisboa, Portugal}
\author{M. Tasinkevych}
\email[]{miko@is.mpg.de}
\affiliation{Max-Planck-Institut f\"{u}r Intelligente Systeme,
             Heisenbergstr. 3, D-70569 Stuttgart, Germany}
\affiliation{IV. Institut f\"ur Theoretische Physik, Universit\"{a}t Stuttgart, Pfaffenwaldring 57, D-70569 Stuttgart, Germany}

\date{\today}

\begin{abstract}
  The Landau-de Gennes free energy is used to study theoretically the 
  effective interaction of a spherical ``key'' and an anisotropic ``lock'' colloidal particles. We assume identical 
  anchoring properties of the surfaces of the key and of the lock particles, and consider planar degenerate and 
  perpendicular anchoring conditions separately. {The lock particle is modeled} as a spherical particle with a spherical dimple. When such a particle is introduced into a  nematic liquid crystal, it orients its dimple at an oblique angle $\theta_{eq}$ with respect to the far field director ${\bf n}_{\infty}$. This angle depends on the depth of the dimple. Minimization results show that the free energy of a pair of key and  lock particles exhibits a global minimum for the configuration when the key particle is facing the dimple of the lock colloidal particle. The preferred orientation $\phi_{eq}$ of the key-lock composite doublet relative to  ${\bf n}_{\infty}$  is robust against thermal fluctuations.  The preferred orientation $\theta_{eq}^{(2)}$ of the dimple particle in the doublet is different from the isolated situation. {This is related to the ``direct'' interaction of defects accompanying the key particle with the edge of the dimple}.  We propose that this nematic-amplified key-lock interaction can play an important role in self-organization and clustering of mixtures of colloidal particles with dimple colloids present.
\end{abstract}

\maketitle

\section{Introduction}

The self-assembly of colloidal particles is an important route towards the design and fabrication of advanced materials with sub-micrometer structures exhibiting new mechanic, electric, and magnetic behavior, and with properties that are essential to relevant applications such as energy harvesting and storage, photonics, electronics, among others \citep{Cademartiri.2015}. However, the production of desired structures with specific symmetries and controlled thermodynamic stability is still an immense challenge. 

In comparison with conventional isotropic host fluids, using liquid crystals (LCs) as dispersing media provides a range of unique opportunities for non-contact colloidal manipulations and ultimately for controlling colloidal assembly. Additionally, the LC host gives rise to long-range anisotropic colloidal interactions that are typically of the order of several $100$ k$_\mathtt{B}$T for sub-micrometer sized particles \citep{Stark.2001}. The mesoscale self-assembly of sub-micrometer particles in LCs may enable composites with properties that can be tuned through a range of external stimuli. Such systems may provide new, cheaper, and more efficient technology, and a fertile ground for basic science \citep{Liu.2014}. For example, due to the mechanical and electro-optical properties of the host, LC colloids are promising candidates for responsive devices -- components that respond optically to applied stress --, or as waveguides to modulate and switch photonic signals \citep{Zografopoulos.2012}.

Recent experimental schemes allow for the semi-assisted assembly of three-dimensional (3D) colloidal structures in LCs that constitute a new class of electrically responsive soft materials \citep{Nych.2013}. Such 3D structures have the advantage over conventional colloidal crystals of combining the mechanical and electro-optical response of LCs to control the inter-particle spacing, thus enabling the fine tuning of photonic band-gaps. However, the current techniques in LC colloids require excessive human intervention and are impractical for large scale production. One way of reducing excessive human intervention during the assembly process is through the use of templates, which can attract and strongly localize colloidal particles  \citep{Silvestre.2004,Hung.2007,Cheung.2008,Silvestre.2014,Eskandari.2014}. Another alternative is to explore how shape anisotropy affects the interaction between colloidal particles, which we address in this manuscript.

Current technology allows for the synthesis of particles mono-dispersed in size, and with prescribed shape that in turn may be used to tune particle interactions \citep{Sacanna.2011,Sacanna.2013,Matijevic.2011,Zhang.2011}. Shape is one of the essential properties of colloids and current techniques make it possible {to synthesize particles with the shape of} ellipsoids, cubes, and prisms, {or} more complex ones such as dumbbells, dimple particles, as well as particles of non-trivial {surface} topology \citep{Sacanna.2011, Lee.2011, Senyuk.2013, Liu.2013, Martinez.2014, Martinez.2015}. Shape-anisotropic particles enable the design of shape-selective interactions and hierarchical self-assembly schemes, and are important in a wide range of applications such as biosensing, diagnostics, and targeted drug delivery. The size of such particles ranges from nano- to micrometer scales, which makes them ideal building blocks for a new generation of responsive materials with controllable electro-optical properties \citep{Glotzer.2007,Frenkel.2011}.

In LC hosts, micrometer particles with regular shapes \citep{Dontabhaktuni.2014} or with non-trivial topology \citep{Senyuk.2013} have been used to explore topological concepts in physical systems, and particles whose surface features a strong curvature have been used to {study the shape-induced localization} of topological defects and the {ensuing} inter-particle interactions \citep{Beller.2015}. {However, the understanding of how shape complementarity may drive the hierarchical assembly of colloidal particles in LC hosts is still lacking. Such knowledge will pave the way to the development of novel colloidal structures with flexible bonds and binding strengths tunable by, e.g., temperature or external fields}. 

Dimple particles are a type of particles that bind to others through shape complementarity \citep{Sacanna.2010,Sacanna.2011,Sacanna.2013}. They are obtained from biphasic colloids whose phases can be selectively and independently polymerized, liquefied, and dissolved \citep{Sacanna.2013}. Typically a polystyrene solid particle will serve as a seed to nucleate the dimple particle; the shape and size of the dimple/cavity depends on the polystyrene particle {size}, which is later dissolved (see Fig. \ref{dimpleP}). When mixed in solution with particles identical to the polystyrene seeds, dimple particles take the role of {\it locks} and bind with particles that can perfectly fit their cavity, the {\it keys} \citep{Sacanna.2010}. {The main mechanism for binding, in the case of 
an isotropic host, is due to entropic forces. Orientational degrees of freedom of a LC host will result in additional long-range anisotropic effective
forces mediated by the distortions of the LC director field.}
 { Moreover, previous studies {of} the LC-mediated interactions of a colloidal particle with a cavity, or a channel, on a planar substrate \citep{Silvestre.2004,Hung.2007,Cheung.2008,Eskandari.2014} revealed a strong key-lock type binding. }

In this manuscript we consider the inclusion of dimple particles in a nematic LC host. The system is model by the Landau-de Gennes free energy as described in the next section. As the shape anisotropy of the particle affects the distribution of topological defects nucleated in the LC matrix, we begin by addressing the problem of a single isolated dimple particle in Section \ref{single}. 
We will show that the dimple adds a region of strong curvature to the colloidal particle, pinning the defect and forcing the particle to orient the dimple in a preferred direction. 
{In Section \ref{key-lock} we explore the LC-mediated interactions between the key and the lock}. Particularly we show that the dimple induces a strong {attraction} and that the binding process is accompanied by a transformation of the topological defects.

\section{Model}

The nematic LC is {modeled by the Landau-de Gennes free energy functional \citep{Gennes.1993},
$F = \int_\Omega{dv\left(f_{b}+f_{e}\right)} + \int_{\partial\Omega}{ds\, f_s}$, written in invariant terms of the tensor order parameter $\Qij$, where

\begin{eqnarray}
f_b &=& a_o \left(T-T^*\right) \Qij^2 - b Q_{ij}Q_{jk}Q_{ki} + c\left(\Qij^2\right)^2 \\
f_e &=& \frac{L}{2}\left(\partial_k \Qij\right)^2.
\end{eqnarray}
$Q_{ij}$ is a traceless symmetric $3\times3$ matrix, and summation over repeated indices is implied. $a_o$, $b$, $c$, and $L$ are phenomenological material-dependent parameters, $T$ is a the temperature, and $T^*$ is the supercooling temperature of the isotropic phase. For simplicity we assume the one-elastic-constant approximation.
Homeotropic surfaces are taken into account by the Nobili-Durand potential \cite{Nobili.1992}
\begin{equation}
f_s^\perp=W\left(\Qij-\Qij^s\right)^2,
\end{equation}
where $W$ is the anchoring strength, and $\Qij^s$ is the surface preferred order parameter. We assume that surfaces with planar anchoring are degenerate and thus are described by the Fournier-Galatola surface potential \citep{Fournier.2005}
\begin{equation}
f_s^\| = W\left(\left(\tilde{\Qij}-\tilde{\Qij}^\perp\right) +  \left(\tilde{\Qij}^2 - \frac{3}{2}Q_b^2\right)^2\right),
\end{equation}
where $\tilde{\Qij} = \Qij + Q_b\delta_{ij}/2$, $\tilde{\Qij}^\perp$ is the projection of $\tilde{\Qij}$ onto the confining surface, and $Q_b$ is the bulk value of the scalar order parameter.

We use the following values for the model parameters, corresponding to typical values for the 5CB \citep{Kralj.1991} $a_o=0.044$ MJ/Km$^3$, $b=0.816$MJ/m$^3$, $c=0.45$ MJ/m$^3$, $T^* = 307$ K, $T=307.2$ K , and $L= 6$ pJ/m.

The total free energy $F$ is minimized numerically using finite elements methods with adaptive meshing, described in detail in \citep{Tasinkevych.2012}.

\section{Results}

\begin{figure}[t]
\centering
\includegraphics[width=0.9\columnwidth]{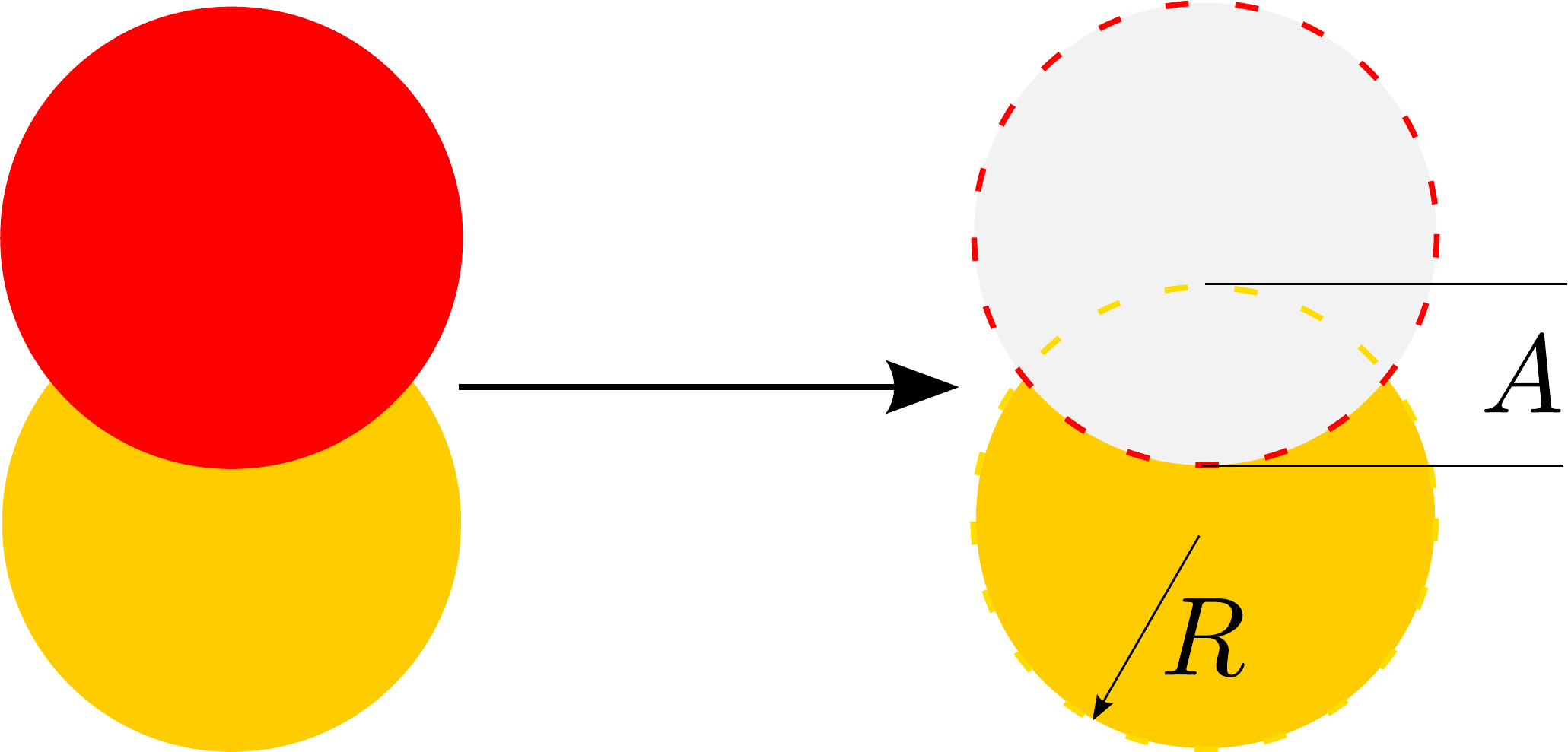}
\caption{ (Color online) Sketch of the formation of a dimple particle. The red particle is the seed. The yellow particle, of radius $R$, corresponds to the dimple particle. When removing the volume occupied by the red particle the yellow particle is left with a dimple of depth $A$, measured from the original spherical surface. }
\label{dimpleP} 
\end{figure}

\begin{figure}[t]
\centering
\includegraphics[width=0.45\columnwidth]{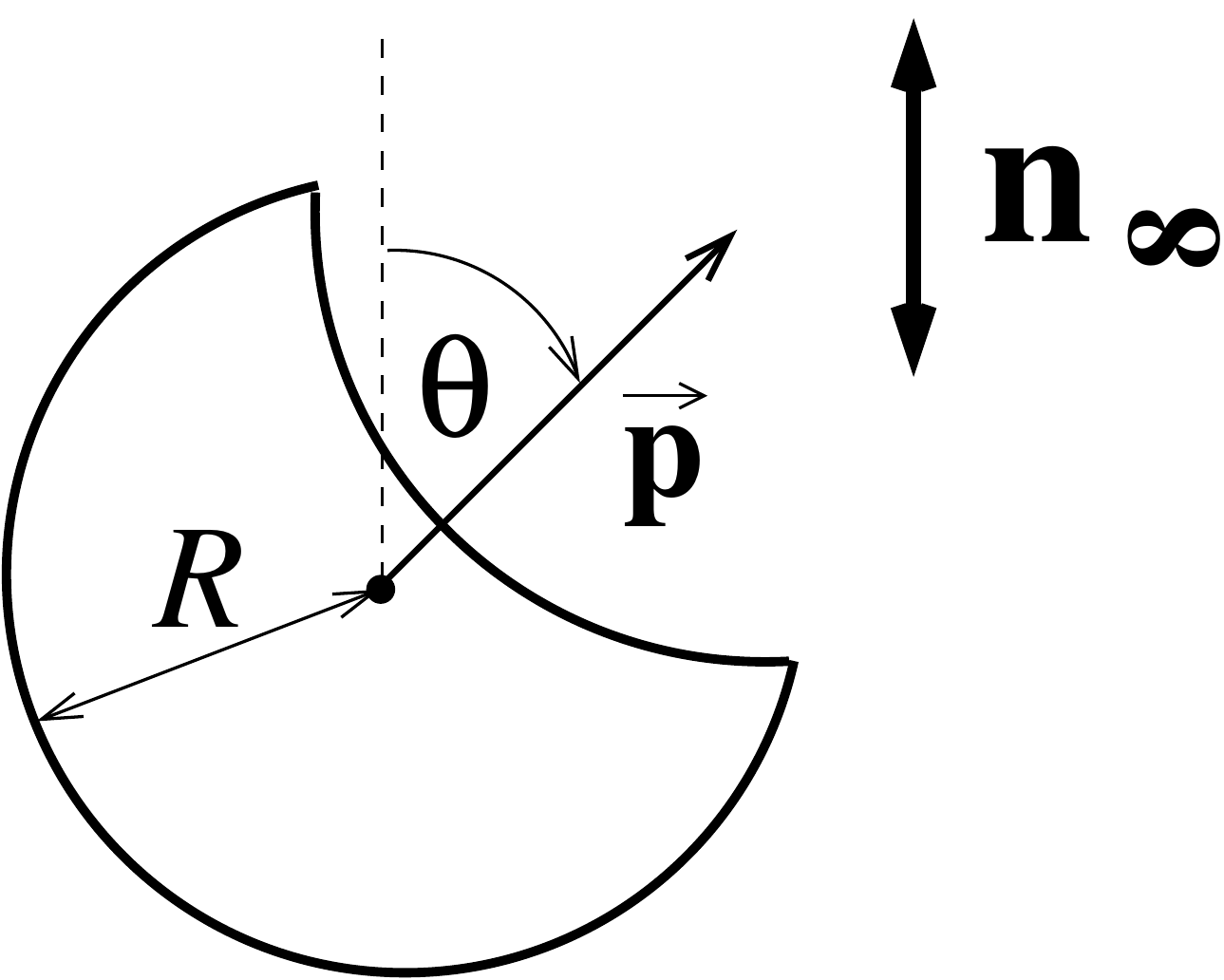}
\caption{Schematic representation of a dimple colloidal particle with the radius $R$. The orientation of the particle is described by
the vector ${\bf p}$ which makes an angle $\theta$ with respect to the far field director ${\bf  n}_{\infty}$. }
\label{dimpleP_geometry} 
\end{figure}

\begin{figure*}
\centering
\includegraphics[width=2.0\columnwidth]{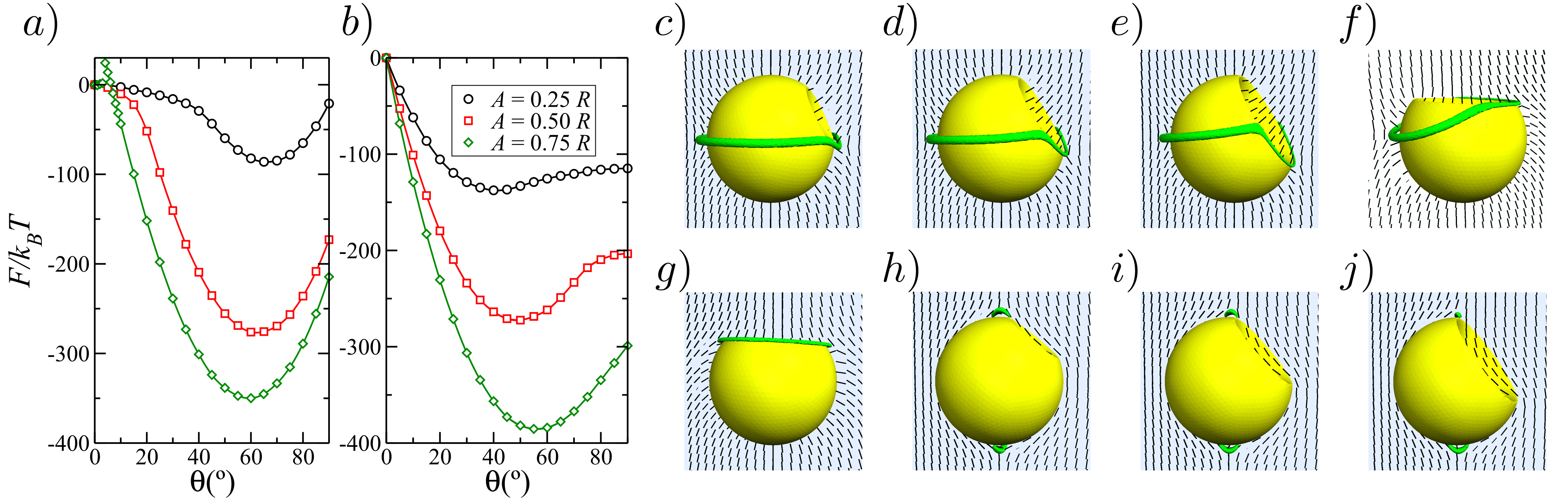}
\caption{(Color online) Landau-de Gennes free energy of an isolated dimple particle as a function of the orientation of the dimple relative to the far field director  $\nvec_\infty$, for several dimple depths $A/R=0.25,\,0.5,\,0.75$. a) Free energy for a dimple particle with homeotropic anchoring. b) Free energy for a dimple particle with planar degenerate anchoring. The reference state is taken for $\theta=0^\circ$. The panels represent the nematic configurations around dimple particles close to their equilibrium orientation $\theta_{eq}$. c) -- e) For homeotropic anchoring $60^\circ \lesssim \theta_{eq}\lesssim 66^\circ$. h)--j) For planar degenerate anchoring $40\lesssim\theta_{eq}\lesssim 56^\circ$. The dimples have depths c) and h) $A=0.25 R$, d) and i) $A=0.5 R$ , e)--g) and j) $A=0.75 R$. For homeotropic anchoring, particles with large dimples exhibit a local minimum at $\theta=0^\circ$. f) Configuration of homeotropic dimple particle with $A=0.75R$ oriented at $\theta=4^\circ$. g) Metastable configuration of homeotropic dimple particle with $A=0.75R$ for $\theta=3^\circ$. The topological defects are represented as green isosurfaces of the scalar order parameter. The black bars represent the director field $\nvec$. $R=0.1\,\mu$m}
\label{singleDimple} 
\end{figure*}

\begin{figure}[t]
\centering
\includegraphics[width=0.7\columnwidth]{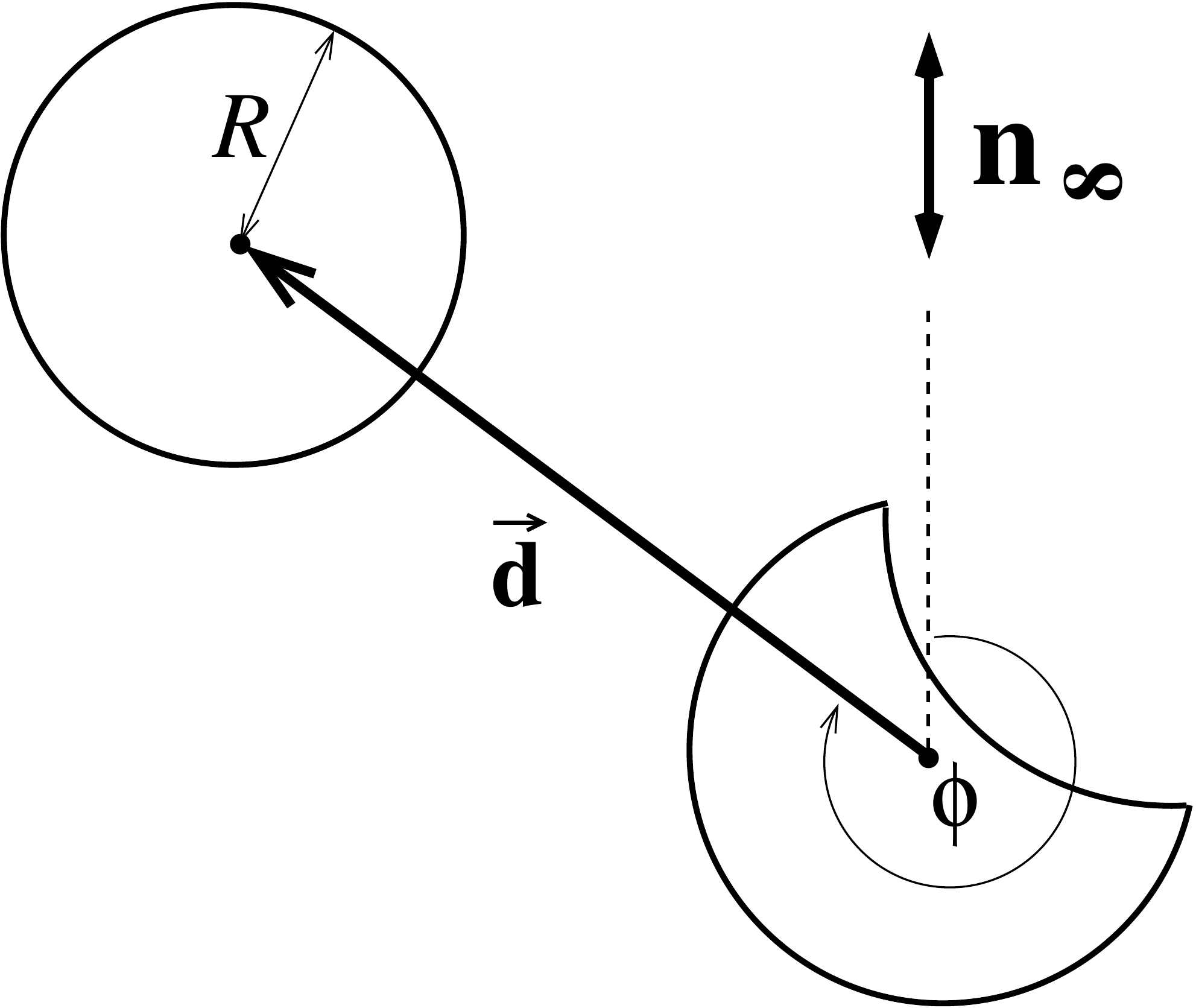}
\caption{Schematic illustration of a pair of a dimple {\it lock} particle and a spherical {\it key} particle, both with the radius $R$.
The center-to-center vector ${\bf d}$ makes an angle $\phi$ with respect to the far field director ${\bf  n}_{\infty}$. Upon 
varying ${\bf d}$ and $\phi$, the orientation of the dimple particle is kept fixed.}
\label{2particle_geometry} 
\end{figure}

\begin{figure*}[t]
\centering
\includegraphics[width=1.75\columnwidth]{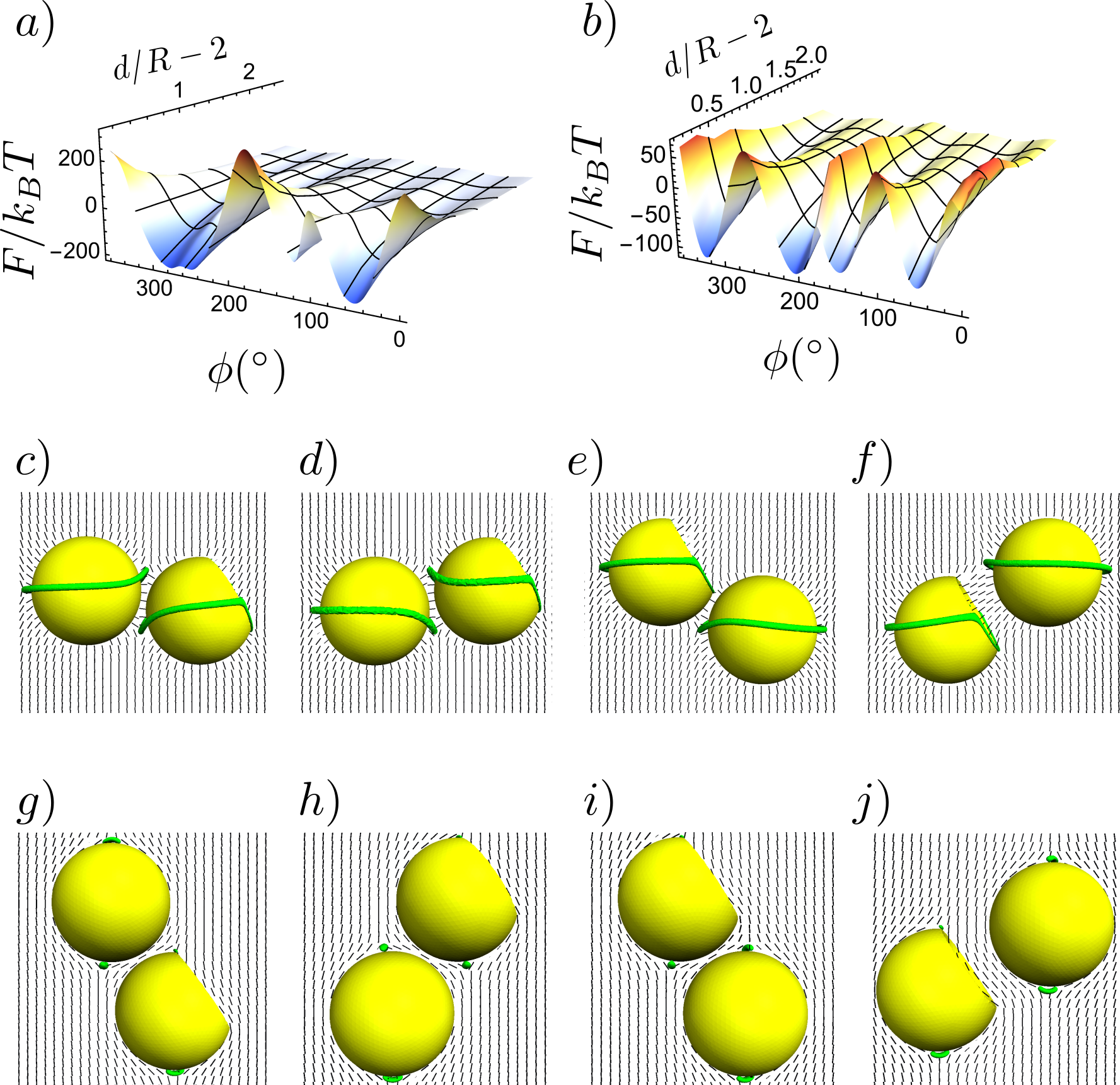}
\caption{(Color online) Landau-de Gennes free energy $F$ for a  pair of a key and a lock particles as a function of the angle $\phi$ between the far field director ${\bf  n}_{\infty}$ and the center-to-center vector ${\bf d}$, and the  distance $d$ between particles, see Fig.~\ref{2particle_geometry}.  a) Homeotropic; and b) planar degenerate anchoring conditions. In both cases the orientation of the 
dimpled particle is fixed at $\theta = \theta_{eq}$, see Fig.~\ref{singleDimple}. Here $2.1R\leqslant d\leqslant 4.5R$. At $d=2.1R$, the free energy profiles exhibits four minima and the corresponding nematic configurations are show in the panels c) $\phi=280^\circ$, d) $\phi=258^\circ$, e) $\phi=120^\circ$, and f) $\phi=60^\circ$ for homeotropic anchoring; and in g) $\phi=330^\circ$, h) $\phi=210^\circ$, i) $\phi=150^\circ$, and j) $\phi=56^\circ$ for planar degenerate anchoring. The topological defects are show as green isosurfaces of the scalar order parameter $Q$ and the director field $\nvec$ is represented by the black bars. Particles have equal radii, $R=0.1\,\mu$m.} 
\label{double} 
\end{figure*}

\begin{figure}[t]
\centering
\includegraphics[width=1.0\columnwidth]{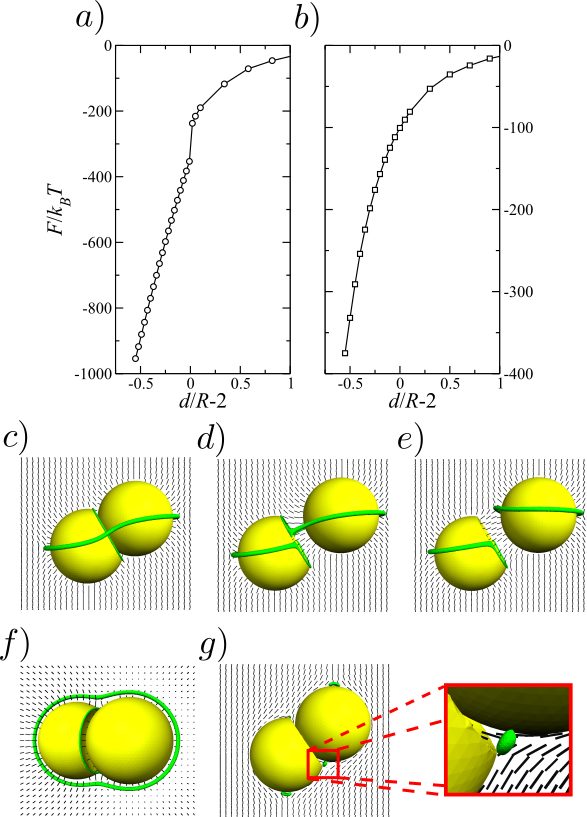}
\caption{{(Color online) Landau-de Gennes free energy $F$ for a  pair of a  {\it key} and a {\it lock} particles as a function of the distance $d$ at $A=0.75R$. The particles have equal radii, $R=0.1\,\mu$m. The orientation of the dimple particle is fixed at $\theta = \theta_{eq}$ ($\theta_{eq}\simeq 60^{\circ}$ for homeotropic anchoring, and $\theta_{eq}\simeq 56^{\circ} $ for planar one, see Fig.~\ref{singleDimple}). a) Homeotropic anchoring, $\phi = 60^{\circ}$; b) planar degenerate anchoring, $\phi = 56^{\circ}$. c) and d) Nematic configuration corresponding to the left branch of $F$ in (a) at distances c) $d=1.45R$ and d) $d=1.99R$. e)  Nematic configuration corresponding to the right branch of the free  energy in (a) at $d=2.02R$. f) Alternative view of the entangled colloids showing how the ring defect is detached from the edge of the dimple. g) Colloidal particles with planar degenerate anchoring at $d=1.45R$. The inset shows the bottom boojum of the {\it key} particle occupying the region of high nematic deformation produced by the edge of the dimple. The topological defects are shown as green isosurfaces of the order parameter $Q$, and the black bars represent the director field $\nvec$.}}
\label{lock} 
\end{figure}

\begin{figure}[t]
\centering
\includegraphics[width=0.9\columnwidth]{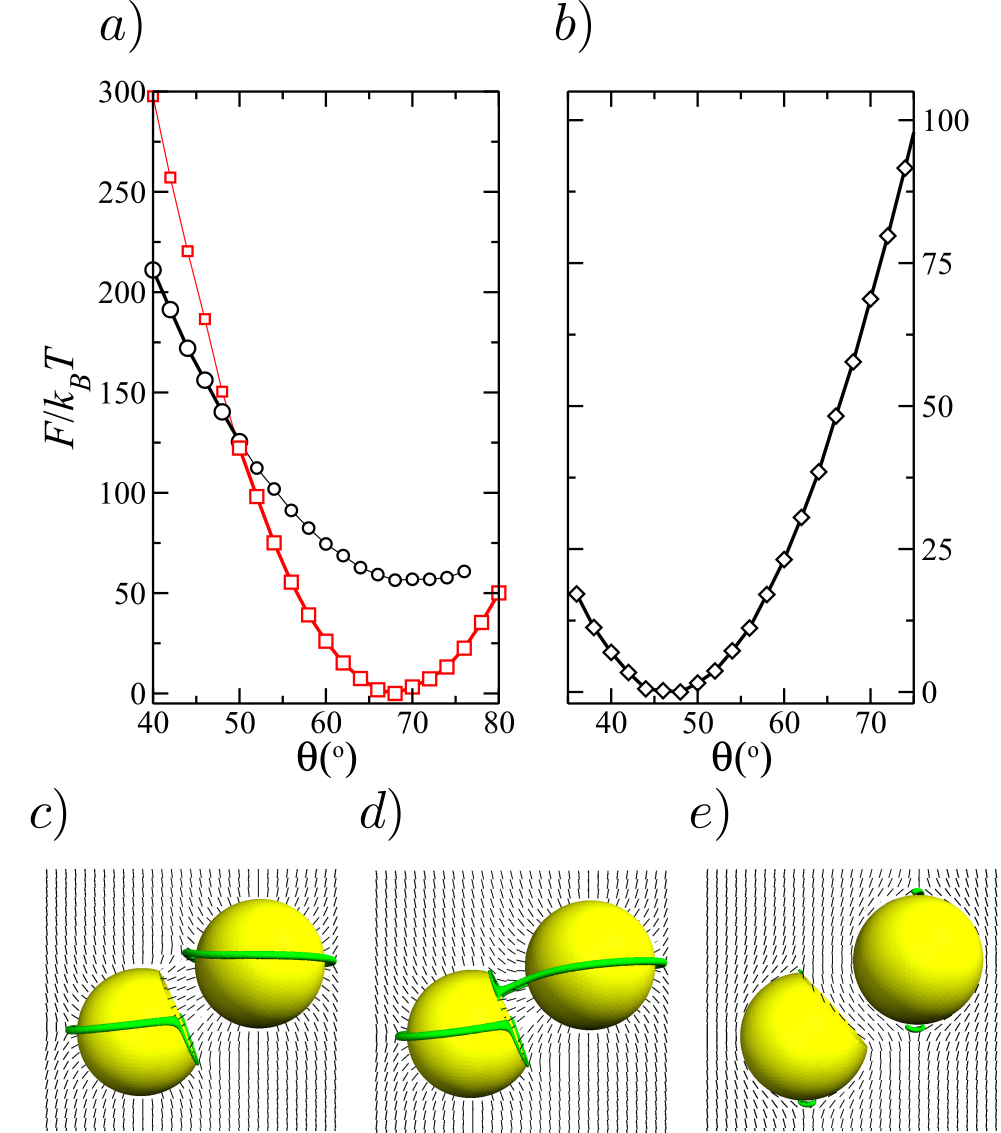}
\caption{(Color online) Landau-de Gennes free energy $F$ as a function of the orientation $\theta$ of the dimple particle (see Fig.~\ref{dimpleP_geometry}), for a) homeotropic and b) planar degenerate anchoring conditions. In both cases $d=2.1R$, and {$\phi=\phi_{eq}\simeq 60^\circ$ for homeotropic anchoring and $\phi=\phi_{eq}\approx 56^\circ$} for planar degenerate anchoring conditions. $d$ and $\phi$ are defined as in Fig.\ref{double}. {We take as a reference the local minimum of the free energy $F(\theta=\theta_{eq}^{(2)})$, where $\theta_{eq}^{(2)}\simeq 68^\circ$ for homeotropic anchoring and $\theta_{eq}^{(2)}\simeq 48^\circ$ for planar degenerate anchoring conditions}. The particles have equal radii, $R=0.1\,\mu$m. The panels show the nematic configuration around homeotropic colloidal particles for dimple orientation $\theta=68^\circ$ for c) the black (circles) energy branch and for d) the red (squares) energy branch shown in (a). e) nematic configuration around particles with planar anchoring for dimple orientation $\theta=48^\circ$. The topological defects are shown as green isosurfaces of the order parameter $Q$, and the black bars represent the director field $\nvec$.}
\label{torque} 
\end{figure}

Dimple particles can be produced either by controlled shell buckling of polymerized silicon oil droplets \citep{Sacanna.2010} or by nucleation growth on a polystyrene seed particle \citep{Sacanna.2013}. The latter has the advantage of obtaining a dimple/cavity with a prescribed shape and with controllable size. Inspired {by} this particular technique we consider that our dimple particles are obtained from the {intersection} of two spherical particles, as depicted in Fig.~\ref{dimpleP}. {The resulting dimple particle, similarly to the experiments \citep{Sacanna.2013}, exhibits a 
sharp edge} and is described by three parameters: (i) the radius $R$ of the spherical section, (ii) the depth of the dimple $A$ measured from the original spherical surface of the particle, and (iii) the radius of the seed particle. {For simplicity, we consider the key and lock particles of equal radii, $R$}. We also assign to the dimple particle the unit vector ${\bf p}$, see Fig.~\ref{dimpleP_geometry}, which describes its orientation.}

\subsection{Single lock particle}
\label{single}

 Usually, colloidal particles align LC molecules either perpendicular (homeotropic anchoring) or parallel (planar degenerate anchoring) to their surfaces \citep{Poulin.1998}. Recently it was reported that polystyrene colloidal particles   {can align} the LC molecules at an oblique angle (conical degenerate anchoring) giving rise to a new type of LC colloids \cite{Senyuk.2016}. Here we restrict our study to homeotropic and planar degenerate anchorings. 

When a colloidal particle is introduced in a uniformly aligned nematic LC, the preferred alignment on its surface introduces an orientational frustration that leads to the nucleation of topological defects. For {the case of} homeotropic anchoring two types of topological defects may appear near the particle: (i) a point-like defect and (ii) a ring defect. While the former appears mainly for large colloidal particles $R>1 \,\mu$m, the latter is nucleated for smaller {sizes} or under strong confinement. {Here we set $R=0.1\,\mu$m and hence, for the case of homeotropic anchoring, the nucleation of ring defects is expected}.
 {For the case of} planar degenerate anchoring, two surface defects, known as {\it boojums}, are nucleated at antipodal points. The structure of such defects has been studied in detail \citep{Tasinkevych.2012}.

Typically, surface regions of high curvature induce more elastic deformation than low curvature surfaces. Additionally, since defects are by themselves a product of strong elastic deformation, due to the orientational frustration, they will tend to migrate, whenever possible, to the  { regions of high curvature}. {This reduces the  free energy of the system because of the overlap of the nematic distortions}. 

When a homeotropic dimple particle is introduced in a uniformly aligned nematic LC  {it induces} the nucleation of a ring defect {encircling} the colloidal particle. As the  {edge} of the dimple corresponds to a region of high curvature, the defect {line deforms in an attempt to migrate towards the edge of the dimple. The extent of the deformation of the defect line depends upon the angle $\theta$ between the vector that defines the orientation of the dimple, ${\bf p}$, and the global orientation $\nvec_\infty$ (see Fig.~\ref{dimpleP_geometry}).}  {Therefore, the first question we address here is \textit{what is the orientation $\theta_{eq}$ that the dimple particle will assume once it {is allowed} to rotate}?} 

In Fig.~\ref{singleDimple}a we show the free energy of an isolated dimple particle as a function of  {$\theta$} for {several values of $A$}. For comparison we take $F(\theta=0^\circ)$ as the reference state, for which the ring defect is axially symmetric and is located close to the particle equator. Our  {numerical results} show that there is a clear preferred orientation  {$\theta_{eq}$} that varies from {$\simeq 60^\circ$ at $A=0.75R$, to $\simeq 66^\circ$ at $A=0.25R$}.  {The depth of the corresponding free energy well is of the order of several $100$ k$_B$T}. In Figs.~\ref{singleDimple}c, \ref{singleDimple}d, and \ref{singleDimple}e we show  {the equilibrium nematic configurations around dimple particles, with homeotropic anchoring, for $A/R=0.25,\,0.5,\,0.75$, respectively.}  {The ring defect is undeformed as long as it goes around the spherical region of the particle, but as the ring reaches the vicinity of the dimple {it bends downwards} and partially circumvents the lower portion of the edge. We have found that for $A=0.75R$, homeotropic dimple particles exhibit a metastable state at $\theta=0^\circ$. As shown in Fig.\ref{singleDimple}f for $\theta=4^\circ$, if the particle is rotated from $\theta_{eq}$ to $\theta=0$ the unpinned section of the ring defect approaches the edge of the dimple. For dimple orientations closer to $\theta=0$ the ring is fully pinned to the edge of the dimple (see Fig.\ref{singleDimple}g for $\theta=3^\circ$).

 When the dimple particle has planar degenerate anchoring two boojums are nucleated at its surface. In this case one of the defects will attempt to migrate to the vicinity of the dimple edge, thus inducing the  {rotation of the particle} towards some preferred orientation. In Fig.~\ref{singleDimple}b we plot the free energy of such particles as a function of   {$\theta$}, for several values of $A$. Again, we take $F(\theta=0^\circ)$ as the reference state. In this case, one of the boojum defects  prefers to sit at the  {edge of the dimple}, as can be seen in Figs.~\ref{singleDimple}h, \ref{singleDimple}i, and \ref{singleDimple}j. As a result the particle preferred orientation $\theta_{eq}$ varies from $\simeq 40^\circ$ at $A=0.25R$ to $\simeq 56^\circ$ at $A=0.75R$. We stress that in both anchoring cases the particles are aligned through strong elastic forces.

\subsection{Lock and key particles}
\label{key-lock}

{We now consider LC-mediated interactions between a dimple ({\it lock})  with a spherical ({\it key}) particles, both having identical anchoring properties, which can perfectly fit   {into} the dimple, i.e., we assume that the radii of both the dimple and of the spherical particle are equal. {In the case of non-identical anchoring conditions we expect a repulsion of the spherical key from the dimple, as reported for the interaction of a colloidal particle with a cavity patterned on a planar surface \citep{Eskandari.2014}.}

The {behavior of spherical colloidal particles dispersed in a LC medium is well studied}  \cite{Lubensky.1998,Gu.2000,Pergamenshchik.2010,Poulin.1998, Smalyukh.2005,Tasinkevych.2012}. At large distances, $d$, two sub-micrometer particles with homeotropic (planar degenerate) anchoring exhibit an effective pair-potential with the symmetry of a quadrupole-quadrupole interaction {decaying as} $d^{-5}$. {The particles repel each other along the directions parallel or perpendicular to the far field director $\nvec_\infty$, and attract when the center-to-center vector is aligned at oblique angles $\simeq 49^{\circ}, 131^{\circ}$ to $\nvec_\infty$ \cite{Lubensky.1998}}. At short distances the interaction is dominated by the nonlinear contributions from the defects nucleated by each particle. As a result the particles bind at oblique angles that are different from the long range ones \cite{Tasinkevych.2012}. Particles with homeotropic anchoring also exhibit a short distance repulsion that prevents them  from touching. This repulsion, however, is not present when the colloidal particles have planar degenerate anchoring \cite{Smalyukh.2005}.

Here we focus on dimple particles with depth $A=0.75R$. The reason being that in this way the dimple can accommodate more surface area from the spherical particle. We have performed the same analysis for dimple particles with $A/R=0.25,\,0.5$ and found qualitatively similar results. The geometry of the system is illustrated in Fig.~\ref{2particle_geometry}.

Figure \ref{double} shows the  {3D landscape} of the  {effective} interaction potential between dimple and spherical particles with homeotropic (Fig.\ref{double}a) and planar degenerate anchoring (Fig \ref{double}b). The  {centers of both} particles are  {restricted to be} on the same plane that bisects the dimple. Additionally, the orientation of the dimple particle is fixed at the single-body equilibrium value $\theta_{eq}$ ($\theta_{eq}\simeq 60^{\circ}$ for homeotropic anchoring, and $\theta_{eq}\simeq 56^{\circ} $ for planar one, see Fig.~\ref{singleDimple})}.  {Under these conditions the free energy landscapes exhibit four troughs along the $d-$direction  at some fixed values of $\phi$ depending on the anchoring type. The three {troughs with  larger $\phi$} are reminiscent of the usual quadrupole-quadrupole interaction of spherical colloids, and the corresponding nematic configurations are shown in Figs.~\ref{double}c to \ref{double}e and Figs.~\ref{double}g to \ref{double}i. {The trough with the smallest value of $\phi=\phi_{eq}$ ($\phi_{eq}\simeq 60^{\circ}$ for homeotropic anchoring, and $\phi_{eq}\simeq 56^{\circ} $ for planar one)} corresponds to the interaction between the spherical particle and the dimple itself, i.e., describes the energetics of the key-and-lock arrangement (Figs.~\ref{double}f and \ref{double}j).} For both anchoring  {types} the  {depth of all four free energy troughs at a distance $d=2.1R$ is of the order of} $100$ k$_B$T. This indicates that depending on the  {initial positioning} of the spherical particle  {relative to the dimple {one}}, the {former} could avoid being attracted towards the dimple. 

To fully assess the strength of the attractive interaction between the {key and the lock}, we have calculated the interaction energy along the {center-to-center orientation} $\phi=\phi_{eq}$,  and at fixed {dimple orientation} $\theta = \theta_{eq}$.
 {The results are} shown in Fig.~\ref{lock} for a) homeotropic {($\phi_{eq}\simeq \theta_{eq}\simeq60^\circ$)} and b) planar degenerate anchorings {($\phi_{eq}\simeq \theta_{eq}\simeq56^\circ$)}. It is clear  {that the configuration ``key-in-the-lock'' corresponds to the global free energy minimum with binding potential of the order of} {$\approx 1000$ k$_B$T for homeotropic and $\approx 300$ k$_B$T for planar degenerate anchoring}. We note that in the case {of homeotropic anchoring} the  {free energy}  {has} two branches that differ by the configuration of the topological defects. Figures \ref{lock}c and \ref{lock}d, and \ref{lock}e illustrate such configurations for the lower  and upper  branches, respectively.  {While in the upper branch each colloidal particle is surrounded by its own defect line}, it turns out that in the binding process the ring defect of the {\it key} particle attaches to the edge of the dimple inducing the {individual} topological defects to merge together forming an entangled dimer. The colloids are embraced by a single defect ring  {which is now detached from the edge of the dimple}, as  {is shown} in Fig.~\ref{lock}f.  {This defect rearrangement} leads to a stronger binding force,  as is seen from the abrupt change in slope of the interaction energy in Fig.~\ref{lock}a. 

In the case of planar degenerate anchoring, the  {free energy varies continuously with $d$ as the key enters into the lock}, Fig.~\ref{lock}d. However, careful inspection of the LC configuration when the spherical colloidal particle has entered the dimple reveals that here too there is  {a rearrangement of the surface topological defects}, {as it is shown in Fig.~\ref{lock}g}. A zoom to the lower region of the spherical particle reveals that 'southern' topological defect approaches the edge of the dimple occupying a region of strong nematic deformation (see the inset of Fig.~\ref{lock}g).

{Finally, we address the ``stiffness'' of the key-lock particles' alignment, {$\phi=\phi_{eq}$}, when dimple is allowed to deviated from the single-particle preferred orientation $\theta_{eq}$. Figure~\ref{torque} shows the free energy of the key-lock pair as a function of the orientation of the dimple particle $\theta$ at $d=2.1R$. We take $F(\theta=\theta_{eq}^{(2)})$ as the reference state, where $\theta_{eq}^{(2)}$ corresponds to the minimum value of $F$. {For both anchoring cases} the dimple particle wants to adjust its {orientation} in such a way that it ``captures'' the spherical particle. However, the angle that the dimple particle adopts when the spherical particle is close depends also on the mutual attraction between the edge of the dimple -- where strong nematic deformations occur --, and the defect of the {\it key} particle. 

The interaction free energy of homeotropic particles exhibits two branches, shown in Fig.\ref{torque}a. We have highlighted the branch with minimum energy by enlarging the symbols and the thickness of the joining lines. The branch in black (circles) corresponds to the configuration shown in Fig.\ref{torque}c, while the branch in red (squares) corresponds to the configuration in Fig-\ref{torque}d. Both configurations are taken for $\theta=\theta_{eq}^{(2)}\simeq68^\circ$. In the first branch the {\it key} and {\it lock} pair are in a disjoint state. In contrast, the second branch corresponds to a state where the ring defect of the spherical particle is deformed and occupies a section at the edge of the dimple. These indicate that at close distance there is a strong interaction between the defect of the {\it key} particle and the edge of the dimple, as a mechanism to reduce the regions of strong LC deformations. For $\theta<50^\circ$, stretching the ring defect is energetically more costly and the system prefers to be in a disjoint state, Fig.\ref{torque}c. However, the dimple is lead to rotate in such a way that its edge approaches the {\it key} particle, thus promoting the ring defect of the latter to stretch towards the edge of the dimple, Fig.\ref{torque}d. The disjoint state is in fact metastable but, as was seen in Fig.\ref{lock}, there is a radial attractive force that brings the particles together promoting the joint state that develops into the entangled state of Figs.\ref{lock}c and \ref{lock}f.

Figure \ref{torque}b shows the interaction energy for the particles with planar anchoring, under the same constraints. As the surface defects are not as large as the ring defects, the system is always in a disjoint state and the free energy is smooth. Nonetheless, also in this case the dimple is deviated from its isolated preferred orientation $\theta_{eq}\simeq56^\circ$ as the edge of the dimple approaches the 'southern' boojum of the {\it key} particle, thus adopting a new equilibrium orientation at $\theta_{eq}^{(2)}\simeq48^\circ$.

\section{Conclusions}

{Within the Landau-de Gennes formalism we have studied the behavior of a single dimple particle, and a pair of a dimple and spherical particles, in a nematic liquid crystal. The dimpled particles are constructed by superimposing of two spherical particles, with shifted centers, and removing the volume occupied by one of them. The resulting ``left-over'' particle has a spherical dimple. We have restricted our attention only to the case when these spheres are of equal sizes, but varying their center-to-center distance. We have considered two types of anchoring conditions at the surfaces of colloidal particles: homeotropic and planar degenerate ones.}

{ We have found that an isolated dimple particle aligns the dimple {at a} specific angle $\theta_{eq}\neq0$ with respect to the far field director $\nvec_\infty$.  $\theta_{eq}$ depends on the depth of the dimple $A$. The main physical reason for this alignment is related to the migration of the topological defects to the edge of the dimple where the nematic distortions are large. Similar phenomenon  has been reported previously for cubic, cylindrical, ellipsoidal, and toroidal particles \citep{Beller.2015,Hashemi.2015,Senyuk.2016a}, where the particle-induced topological defects localize at the high-curvature regions.}

{{Next we} have studied the {effective LC-mediated} interaction between a spherical {\it key} and a dimple {\it lock} particles {with} identical anchoring conditions.  We have found that the effective pair potential is strongly repulsive along the directions parallel and perpendicular to the far field director $\nvec_\infty$, and attractive at oblique angles, similarly to the quadrupolar interaction 
of spherical colloids.  The fact that the key-lock effective interaction potential exhibits the shape {similar to that} of spherical particles, indicates that the {key may}  be trapped at one of the local free energy minima which is outside of the dimple of the {\it lock}. However, we have found that the depth of the potential well in the configuration when the {\it key} is facing the {\it lock} is roughly an order of magnitude larger than the depth of the other local equilibrium orientations. 

Our calculations also indicate that the key-lock mechanism is accompanied by a rearrangement of the accompanying topological defects. Particularly, we have found that there is a defect sharing mechanism that is triggered when the spherical particle enters the dimple. In the case of homeotropic anchoring the two rings merge into a single {one} entangling the composite dimmer. In the case of planar degenerate anchoring, the particles share one of the boojum defects.} 

{We have found that the presence of the spherical particle induces a change in the equilibrium orientation of the dimpled one. In particular, the preferred dimple orientation increases (decreases) upon the approach of the spherical particle in the case of homeotropic (planar degenerate) anchoring. We relate this effect to the direct short distance interaction of the topological defects of the {\it key} particle with the edge of the dimple particle.} 

{Finally, we note that the role of dimple particles in the growth of large colloidal structures in a nematic LC host is still unclear. However, it should be possible to tailor the valence of the lock particle in such a way that it can serve as a nucleation site with a predefined symmetry. To that end, future studies should address the three- (or more) particles problem. The number of degrees of freedom involved in such type of studies poses a large challenge.}

\acknowledgments

We acknowledge the financial support of the Portuguese Foundation for Science and Technology (FCT) under the contracts numbers 
UID/FIS/00618/2013 and EXCL/FIS-NAN/0083/2012.

\end{document}